\documentclass[superscriptaddress,groupedaddress,graphics,preprintnumbers,twocolumn,prd]{revtex4}
\usepackage{graphicx}
\usepackage{dcolumn}
\usepackage{bm}
\usepackage{amssymb}
\usepackage{mwe}
\usepackage{comment}
\usepackage{xcolor}
\usepackage{amsmath}
\usepackage{tensor}
\usepackage{hyperref}
\usepackage[capitalise]{cleveref}
\usepackage[normalem]{ulem}
\usepackage{orcidlink}

\crefname{section}{Sec.}{Sections}
\crefname{equation}{Eq.}{Equations}
\crefname{figure}{Fig.}{Figures}

\DeclareMathOperator\Arg{Arg}

\begin{document}

\preprint{CTPU-PTC-23-53}

\title{Electromagnetic field in a cavity induced by gravitational waves}

\author{Danho Ahn\orcidlink{0000-0003-2224-9467}}
\email[]{danho.phy@gmail.com} 
\affiliation{Center for Axion and Precision Physics Research, \href{https://ror.org/00y0zf565}{Institute for Basic Science}, Daejeon 34051, Republic of Korea}

\author{Yeong-Bok Bae\orcidlink{0000-0003-3093-9206}}
\email[]{astrobyb@gmail.com} 
\affiliation{Department of Physics, \href{https://ror.org/01r024a98}{Chung-Ang University}, Seoul 06974, Republic of Korea}
\affiliation{Particle Theory and Cosmology Group, Center for Theoretical Physics of the Universe, \href{https://ror.org/00y0zf565}{Institute for Basic Science}, Daejeon 34051, Republic of Korea}

\author{Sang Hui Im\orcidlink{0000-0003-2418-4871}}
\email[]{imsanghui@ibs.re.kr}
\affiliation{Particle Theory and Cosmology Group, Center for Theoretical Physics of the Universe, \href{https://ror.org/00y0zf565}{Institute for Basic Science}, Daejeon 34051, Republic of Korea}

\author{Chan Park\orcidlink{0000-0002-2692-7520}}
\email[]{iamparkchan@gmail.com}
\affiliation{Particle Theory and Cosmology Group, Center for Theoretical Physics of the Universe, \href{https://ror.org/00y0zf565}{Institute for Basic Science}, Daejeon 34051, Republic of Korea}
\affiliation{Center for the Gravitational-Wave Universe, Astronomy Program, Department of Physics and Astronomy,
\href{https://ror.org/04h9pn542}{Seoul National University}, Seoul 08826, Republic of Korea}
\affiliation{Institute for Gravitational Wave Astronomy, \href{https://ror.org/00hy87220}{Henan Academy of Sciences}, Zhengzhou 450046, Henan, China}
\date{August 28, 2024}

\begin{abstract}
The detection method of gravitational waves (GW) using electromagnetic (EM) cavities has garnered significant attention in recent years. This paper thoroughly examines the analysis for the perturbation of the EM field and raises some issues in the existing literature. Our work demonstrates that the rigidity condition imposed on the material, as provided in the literature, is inappropriate due to its reliance on a gauge-dependent quantity that cannot be controlled experimentally. Instead, we incorporate elasticity into the material and revise the governing equations for the electric field induced by GWs, expressing them solely in terms of gauge-invariant quantities. Applying these equations to cylindrical cavities, we present the GW antenna patterns for the detector.
\end{abstract}

\maketitle

\section{Introduction}

Electromagnetic (EM) fields play a crucial role in the observation of gravitational waves (GWs). Interferometer-type GW detectors utilize lasers \cite{abbottGWTC3CompactBinary2023}, and pulsar timing arrays employ EM pulses from millisecond pulsars to detect GWs \cite{theeptaandinptacollaborationsSecondDataRelease2023,reardonSearchIsotropicGravitationalwave2023,
agazieNANOGrav15Yr2023,xuSearchingNanoHertzStochastic2023}. Recently, the ideas of using high-sensitivity EM cavities for GW detection \cite{hermanDetectingPlanetarymassPrimordial2021,aggarwalChallengesOpportunitiesGravitationalwave2021,berlinDetectingHighfrequencyGravitational2022,domckeNovelSearchHighFrequency2022} have been receiving considerable attention with their applications \cite{domcke2023symmetries,berlinElectromagneticCavitiesMechanical2023,navarro2023study}. The principle underlying this method involves resonating EM fields when the frequency of GWs closely matches the resonant frequency of the EM cavity. This approach has the advantage of being relatively easy to try because sensitive cavity experiments are already being conducted worldwide to search for axion dark matter \cite{andrew2023axion,bartram2021search,backes2021quantum}.

The resonant frequency of an EM cavity is determined by the size of the cavity. For instance, assuming a cavity size of approximately $0.1$ meter would yield a resonant frequency in the GHz range. If we consider binary black holes as the source of GWs, planet-mass binary black holes of about $\sim 10^{-5} M_{\odot}$ would be required to produce GWs at GHz frequencies just before merging \cite{Maggiore:2007ulw}. Such mass black holes are challenging to form from stars \cite{subsolarBH_LVK}; hence, if they exist, they are likely to be primordial black holes formed from the density fluctuations in the early universe. Although direct evidence of black holes of this mass has yet to be found, the study \cite{niikuraConstraintsEarthmassPrimordial2019}, that suggests the possibility of planet-mass black holes through gravitational microlensing events is noteworthy.

The phenomena derived from GWs are described by the perturbation theory. In the process of unfolding that, there is freedom in the choice of gauge. However, that choice cannot create any physical differences. Depending on the gauge choice, there are various paths to reach an identical gauge-invariant quantity. When dealing with gauge-dependent quantities, it is easy to fall into the trap of considering gauge artifacts as physical entities. Therefore, to avoid this, it is preferable to describe the governing equations and imposed physical conditions solely in terms of gauge-invariant quantities. To achieve this, we will introduce appropriate gauge-invariant quantities to describe equations and physical conditions. Moreover, in the process, we will not choose any specific gauge.

In this paper, we extensively investigate the interrelation of EM fields, acoustic oscillations, and GWs to deepen our understanding of the operational principles of the EM cavity. The oscillations of the material and the EM field induced by GWs are expressed through the equations containing only gauge-invariant quantities. The prescribed physical conditions are also presented using gauge-invariant quantities, enabling experimental implementation. From this perspective, we critically review existing studies \cite{berlinDetectingHighfrequencyGravitational2022,domckeNovelSearchHighFrequency2022}, pointing out certain issues. Furthermore, we obtain solutions for gauge-invariant quantities from revised equations and physical conditions. This allows us to comprehend the interactions among the EM field, acoustic oscillations, and GWs, and accurately describe the GW signal measurable from the EM cavity.

Our paper is structured as follows. In \cref{sec:laws_in_curved_spacetime}, we introduce physical laws in curved spacetime necessary for discussing the induced EM field by GWs. \cref{sec:em_cavity} covers the basics of the EM cavity. To provide a helpful pedagogical illustration, \cref{sec:forced_oscillation} presents an example of forced oscillation. In \cref{sec:covariant_perturbation}, we present the covariant perturbation theory, enabling a concise perturbation analysis. Applying this approach, we discuss perturbations of Minkowski spacetime in \cref{sec:perturbation_Minkowski} without choosing any gauge. The perturbations of elasticity are developed in \cref{sec:perturbation_elasticity}, and those of electromagnetism are discussed in \cref{sec:perturbation_EM}. For the inside of the vacuum cavity, we provide the equation for the induced electric field by GWs in \cref{sec:inside_cavity}. \cref{sec:gauges} addresses the inadequacy of the rigid condition given in \cite{berlinDetectingHighfrequencyGravitational2022,domckeNovelSearchHighFrequency2022}. Using corrected equations, we present antenna patterns for the detector in \cref{sec:GW_signal}. 

\section{Preliminary}
    
    \subsection{Physical laws in curved spacetime}
    \label{sec:laws_in_curved_spacetime}

    The indices $a,b,\cdots$ represent abstract indices \cite{wald1984general}, while $\alpha,\beta,\cdots$ denote indices of spacetime components in a basis. We set $c=1$ and $G=1$ (geometrized unit), and $\epsilon_{0}=1/4\pi$ and $\mu_{0}=4\pi$ (Gaussian unit). We introduce the metric signature of $\left(-1,1,1,1\right)$. Consider a globally hyperbolic spacetime $\mathcal{M}$, which is described by Einstein's equations as
    \begin{align}
        G_{ab}=8\pi T_{ab},\label{eq:Einsteins_equations}
    \end{align}
    where $G$ is the Einstein tensor and $T$ is the stress--energy tensor of matters. The contracted Bianchi identities yield
    \begin{align}
        \nabla^{b}T_{ab}=0,\label{eq:contracted_Bianchi_identities}
    \end{align}
    where $\nabla$ is the Levi-Civita connection associated with the spacetime metric $g$. Now, let us consider the unit vector field $u$ for a timelike geodesic congruence without vorticity. Its normalization condition is given by
    \begin{align}
        u\cdot u&=-1,\label{eq:normalization_u}
    \end{align}
    where $\cdot$ denotes the inner product defined by $g$. The spatial metric $\gamma$ and the volume form $\varepsilon$ orthogonal to $u$ are defined as
    \begin{align}
        \gamma_{ab}&\equiv g_{ab}+u_{a}u_{b},
        \\\varepsilon_{abc}&\equiv u^{d}\varepsilon^{\left(4\right)}_{dabc},
    \end{align}
    where $\varepsilon^{\left(4\right)}$ is the spacetime volume form. We introduce the extrinsic curvature defined by
    \begin{align}
        K_{ab}\equiv\nabla_{b}u_{a}.\label{eq:K_definition}
    \end{align}
    This curvature is spatial, as contractions between $u$ and all indices of $K$ vanish, and it is symmetric due to the absence of vorticity. By the Ricci identity, the covariant derivative of $K$ is given by
    \begin{align}
        \nabla_{c}K_{ab}&=\nabla_{b}K_{ac}-u_{d}\tensor{R}{^{d}_{acb}},\label{eq:K_derivative}
    \end{align}
    where $R$ is the Riemenn tensor.

    Let us consider a material with a vacuum cavity, where the motion of the cavity forms a four-dimensional volume $\mathcal{W}$ in $\mathcal{M}$. Introduce the unit vector field $v$ for a timelike congruence with the normalization condition
    \begin{align}
        v\cdot v&=-1,\label{eq:normalization_v}
    \end{align}
    such that its values on the material are four-velocities of the material elements. The spatial metric $\gamma'$ and the volume form $\varepsilon'$ orthogonal to $v$ are given by
    \begin{align}
        \gamma'_{ab}&\equiv g_{ab}+v_{a}v_{b},\label{eq:gamma_prime}
        \\\varepsilon'_{abc}&\equiv v^{d}\varepsilon^{\left(4\right)}_{dabc}.
    \end{align}

    The motions of material elements are influenced not only by spacetime but also by the elastic force arising from the deformation of the material. To analyze the elasticity, we require geometrical quantities known as the material metric $\chi$ and the material volume form $\Omega$, as discussed in \cite{hudelistRelativisticTheoryElastic2023}. These quantities are spatial to $v$ in the sense that
    \begin{align}
        \chi_{ab}v^{b}&=0,\label{eq:chi_orthogonal_to_v}
        \\\Omega_{abc}v^{c}&=0.
    \end{align}
    They are also symmetric along $v$, and $\Omega$ is proportional to $\varepsilon'$ as follows:
    \begin{gather}
        \mathcal{L}_{v}\chi_{ab}=0,\label{eq:chi_symmetry}
        \\\mathcal{L}_{v}\Omega_{abc}=0,
        \\\varepsilon'_{abc}=J\Omega_{abc},
    \end{gather}
    where $\mathcal{L}$ is the Lie derivative and $J$ represents the Jacobian.

    The stress--energy tensor of the material is decomposed into
    \begin{align}
        T_{ab}=\rho v_{a}v_{b}+\sigma_{ab},\label{eq:material_stress_energy}
    \end{align}
    where $\rho$ is the energy density and $\sigma$ is the stress. We assume that the material is homogeneous and isotropic in terms of the material metric $\chi$, such that $\rho$ and $\sigma$ satisfy
    \begin{gather}
        \nabla_{a}\left(\rho v^{a}\right)=0,\label{eq:rho_conservation}
        \\\sigma_{ab}=-J^{-1}\left(\lambda\chi_{ab}\left(e_{cd}\chi^{cd}\right)+2\mu e_{ab}\right),
    \end{gather}
    where $\lambda$ and $\mu$ are constant Lamé parameters, and $e$ is the strain defined by
    \begin{align}
        e_{ab}\equiv\frac{1}{2}\left(\gamma'_{ab}-\chi_{ab}\right).\label{eq:strain}
    \end{align}

    We introduce the orthogonal decomposition \cite{Gourgoulhon2012} with respect to $u$. Its derived quantities will be useful in the analysis of perturbation. The orthogonal decomposition of $v$ is given by
    \begin{align}
        v^{a}&=\left(u^{a}+V^{a}\right)\Gamma,
        \\\Gamma&\equiv-\left(u\cdot v\right),\label{eq:Lorentz_factor}
        \\V^{a}&\equiv\left(\gamma v\right)^{a}\Gamma^{-1},\label{eq:spatial_velocity}
    \end{align}
    where $\Gamma$ is the Lorentz factor, $V$ is the spatial velocity, and $\gamma\left(\cdot\right)$ is the projection operator to the tangent subspace orthogonal to $u$ for all indices of the dot. The material metric $\chi_{ab}$ is decomposed into
    \begin{align}
        \chi_{ab}&=\alpha u_{a}u_{b}+u_{a}\beta_{b}+\beta_{a}u_{b}+W_{ab}+\gamma_{ab},\label{eq:chi_decomposition}
        \\\alpha&\equiv\chi_{ab}u^{a}u^{b}=\chi_{ab}V^{a}V^{b},\label{eq:alpha}
        \\\beta_{a}&\equiv-u^{c}\tensor{\gamma}{^{d}_{a}}\chi_{cd}=V^{c}\tensor{\gamma}{^{d}_{a}}\chi_{cd},\label{eq:beta}
        \\W_{ab}&\equiv\left(\gamma\chi\right)_{ab}-\gamma_{ab},
    \end{align}
    where $\alpha$ is the temporal-temporal part, $\beta$ is the temporal-spatial part, and $W$ is taken from the spatial-spatial part of $\chi$. For \cref{eq:alpha,eq:beta}, we utilized \cref{eq:chi_orthogonal_to_v}.

    Electromagnetic fields $F$ are governed by Maxwell's equations as
    \begin{align}
        \nabla^{b}F_{ab}&=4\pi j_{a},\label{eq:Maxwells_eq_1}
        \\dF&=0,\label{eq:Maxwells_eq_2}
    \end{align}
    where $j$ is the EM current and $d$ is the exterior derivative. Using the above, we obtain the wave equation for $F$ as
    \begin{align}
        \Box F_{ab}=-4\pi \left(dj\right)_{ab}-F_{cd}\tensor{R}{^{cd}_{ab}}-2F_{c[a}\tensor{R}{^{cd}_{b]d}},\label{eq:Maxwells_eq_in_wave_eq_form}
    \end{align}
    where $\Box\equiv\nabla^{a}\nabla_{a}$ is the D'Alembertian as Eq.~(6) in \cite{hermanDetectingPlanetarymassPrimordial2021}. The electric field and magnetic fields with respect to the material elements are given by
    \begin{align}
        E_{a}&=F_{ab}v^{b},\label{eq:E_field}
        \\B_{a}&=\frac{1}{2}\tensor{\varepsilon}{^{\prime bc}_{a}}F_{bc},\label{eq:B_field}
    \end{align}
    as pointed out in \cite{hwangDefinitionElectricMagnetic2023}.

    Because the cavity is a vacuum, we set conditions on $\mathcal{W}$ as
    \begin{align}
        \left.V^{a}\right|_{\mathcal{W}}&=0,\label{eq:vacuum_condition_1}
        \\\left.j_{a}\right|_{\mathcal{W}}&=0.\label{eq:j_at_cavity}
    \end{align}
    When the material is a perfect conductor, we can impose the boundary condition on $\partial\mathcal{W}$, which is the three-dimensional timelike hypersurface between the cavity and the material, given by
    \begin{align}
        \left.\tensor{P}{^{b}_{a}}E_{b}\right|_{\partial\mathcal{W}}=0,\label{eq:boundary_condition_E_p}
    \end{align}
    where $\tensor{P}{^{a}_{b}}\equiv\tensor{\gamma}{^{\prime a}_{b}}-n^{a}n_{b}$ is the projection operator, and $n$ is the spacelike unit vector field such that its values on $\partial\mathcal{W}$ are identical to the normal vector of $\partial\mathcal{W}$. Note that $n$ is orthogonal to $v$.
    
    \subsection{EM cavity in Minkowski spacetime}
    \label{sec:em_cavity}

    Consider the EM cavity in Minkowski spacetime, where the material and the EM field are weak enough to ignore changes of Minkowski spacetime. In this case, we can set $u$ as the constant vector field of four-velocity aligning with our laboratory. Inside the cavity, \cref{eq:Maxwells_eq_in_wave_eq_form} by contracting with $u$ becomes the homogeneous wave equation for the electric field. When the material is a perfect conductor, we obtain the stationary solution as
    \begin{align}
        E_{a}\left(t,\vec{x}\right)=2\Re\left[\sum_{n}\tilde{\mathcal{E}}_{n}\mathfrak{e}^{n}_{a}\left(\vec{x}\right)e^{-i\omega_{n} t}\right],\label{eq:E_superposition}
    \end{align}
    where we introduce a globally inertial coordinate system $\left\{t,\vec{x}\right\}$ such that $u=\left(\partial\middle/\partial t\right)$. Here, $\mathfrak{e}^{n}\left(\vec{x}\right)$ is the real resonant mode satisfying the boundary condition \cref{eq:boundary_condition_E_p}, $\omega_{n}$ is the resonant frequency, and $\tilde{\mathcal{E}}_{n}$ is the complex amplitude. The resonant modes satisfy that following properties:
    \begin{align}
        \Delta \mathfrak{e}^{n}_{a}=-\omega_{n}^{2}\mathfrak{e}^{n}_{a},
        \\\int_{\mathcal{V}}d\mathcal{V}\,\mathfrak{e}^{n}\cdot\mathfrak{e}^{m}=\delta^{nm}\mathcal{V},\label{eq:normalization_e}
    \end{align}
    where $\Delta$ is the Laplacian, $\mathcal{V}$ is the spatial volume of the cavity, and $\delta^{nm}$ is the Kronecker delta. Note that $\mathfrak{e}^{n}\left(\vec{x}\right)$ is dimensionless, and $\tilde{\mathcal{E}}_{n}$ has the same dimension as the electric field.

    The resonant modes of a cylindrical cavity are categorized into TM and TE modes. In each category, they have indices $(m,n,p,s)$ where $m=0,1,\cdots$ is the azimuthal and $n=1,2,\cdots$ is the radial mode number. The longitudinal mode number $p$ can be $p=0,1,\cdots$ for TM modes and $p=1,2,\cdots$ for TE modes. For $m\neq0$, we have two degenerated modes denoted with $s=+1,-1$, respectively. The resonant frequencies for TM and TE modes, respectively, are given by
    \begin{align}
        \omega^{\mathrm{TM}}_{mnp}&=\sqrt{\left(\mathfrak{j}_{mn}/R\right)^{2}+\left(p\pi/L\right)^{2}},
        \\\omega^{\mathrm{TE}}_{mnp}&=\sqrt{\left(\mathfrak{j}'_{mn}/R\right)^{2}+\left(p\pi/L\right)^{2}},
    \end{align}
    where $R$ and $L$ are the radius and the length of the cylinder, respectively, $\mathfrak{j}_{mn}$ is the $n$-th zero of the Bessel function $\mathfrak{J}_{m}\left(x\right)$, and $\mathfrak{j}'_{mn}$ is the $n$-th zero of the derivative of Bessel function $\mathfrak{J}'_{m}\left(x\right)$.
    
    In the cylindrical coordinate $\left\{\rho,\phi,z\right\}$ whose origin is located at the center of the cylinder, the resonant modes for TM and TE, respectively, are given by
    \begin{align}
        &\mathfrak{e}^{\mathrm{TM}}_{mnps}\left(\rho,\phi,z\right)=\frac{2\mathfrak{A}^{\mathrm{TM}}_{mnp}}{\mathfrak{J}'_{m}\left(\mathfrak{j}_{mn}\right)}\left[\mathfrak{R}^{\mathrm{TM}}_{mn,z}\left(\rho\right)\mathfrak{C}_{-s}\left(m\phi\right)\mathfrak{Z}_{p+}\left(z\right)\hat{z}\right.
        \nonumber\\&\qquad\left.-\frac{R}{L}\frac{p\pi}{\mathfrak{j}_{mn}}\left\{\mathfrak{R}^{\mathrm{TM}}_{mn,\rho}\left(\rho\right)\mathfrak{C}_{-s}\left(m\phi\right)\hat{\rho}\left(\phi\right)\right.\right.
        \nonumber\\&\qquad\qquad\left.\left.+s\mathfrak{R}^{\mathrm{TM}}_{mn,\phi}\left(\rho\right)\mathfrak{C}_{s}\left(m\phi\right)\hat{\phi}\left(\phi\right)\right\}\mathfrak{Z}_{p-}\left(z\right)\right],
        \\&\mathfrak{e}^{\mathrm{TE}}_{mnps}\left(\rho,\phi,z\right)=\frac{2\mathfrak{A}^{\mathrm{TE}}_{mnp}}{\mathfrak{J}_{m}\left(\mathfrak{j}'_{mn}\right)}\left[s\mathfrak{R}^{\mathrm{TE}}_{mn,\rho}\left(\rho\right)\mathfrak{C}_{s}\left(m\phi\right)\hat{\rho}\left(\phi\right)\right.
        \nonumber\\&\qquad\left.-\mathfrak{R}^{\mathrm{TE}}_{mn,\phi}\left(\rho\right)\mathfrak{C}_{-s}\left(m\phi\right)\hat{\phi}\left(\phi\right)\right]\mathfrak{Z}_{p-}\left(z\right),
    \end{align}
    where $\left\{\hat{\rho}\left(\phi\right),\hat{\phi}\left(\phi\right),\hat{z}\right\}$ is the orthonormal basis of cylindrical coordinate and 
    \begin{align}
        \mathfrak{A}^{\mathrm{TM}}_{mnp}&\equiv\left[\left(1+\delta_{m0}\right)\right.
        \nonumber\\&\times\left.\left\{\left(\frac{R}{L}\frac{p\pi}{\mathfrak{j}_{mn}}\right)^{2}\left(1-\delta_{p0}\right)+\left(1+\delta_{p0}\right)\right\}\right]^{-\frac{1}{2}},
        \\\mathfrak{A}^{\mathrm{TE}}_{mnp}&\equiv\left[1-\left(m/\mathfrak{j}'_{mn}\right)^{2}+\delta_{m0}\right]^{-\frac{1}{2}},
    \end{align}
    \begin{align}
        \mathfrak{R}^{\mathrm{TM}}_{mn,\rho}\left(\rho\right)&\equiv\mathfrak{J}'_{m}\left(\frac{\mathfrak{j}_{mn}}{R}\rho\right),
        \\\mathfrak{R}^{\mathrm{TM}}_{mn,\phi}\left(\rho\right)&\equiv\frac{m}{\mathfrak{j}_{mn}\rho/R}\mathfrak{J}_{m}\left(\frac{\mathfrak{j}_{mn}}{R}\rho\right),
        \\\mathfrak{R}^{\mathrm{TM}}_{mn,z}\left(\rho\right)&\equiv\mathfrak{J}_{m}\left(\frac{\mathfrak{j}_{mn}}{R}\rho\right),
        \\\mathfrak{R}^{\mathrm{TE}}_{mn,\rho}\left(\rho\right)&\equiv\frac{m}{\mathfrak{j}'_{mn}\rho/R}\mathfrak{J}_{m}\left(\frac{\mathfrak{j}'_{mn}}{R}\rho\right),
        \\\mathfrak{R}^{\mathrm{TE}}_{mn,\phi}\left(\rho\right)&\equiv\mathfrak{J}'_{m}\left(\frac{\mathfrak{j}'_{mn}}{R}\rho\right),
    \end{align}
    \begin{align}
        \mathfrak{C}_{s}\left(\phi\right)&\equiv\begin{cases}
            \cos\phi&:s=+1
            \\\sin\phi&:s=-1
        \end{cases},
        \\\mathfrak{Z}_{ps}\left(z\right)&\equiv\mathfrak{C}_{s}\left(\frac{p\pi}{L}\left(z+\frac{L}{2}\right)\right).
    \end{align}
    For $m=0$, we have to choose $s=-1$ to get a nontrivial mode. Note that the above are normalized by \cref{eq:normalization_e}. In addition, these modes have the parity symmetry as
    \begin{align}
        \mathfrak{e}_{mnps}\left(\rho,\pi+\phi,-z\right)=\left(-1\right)^{m+p}\mathfrak{e}_{mnps}\left(\rho,\phi,z\right).\label{eq:e_symmetry}
    \end{align}

    \subsection{Forced oscillation}
    \label{sec:forced_oscillation}
    
    Let us consider a pedagogical example of the resonance of a mass $m$ attached to a spring with coefficient $k$ and subjected to an external force $f\left(t\right)$. The equation of motion is given by
    \begin{align}
        m\ddot{x}+b\dot{x}+kx=f\left(t\right),\label{eq:force_oscillation}
    \end{align}
    where $x\left(t\right)$ is the displacement of the mass from the equilibrium point, and $b$ is the friction coefficient. By applying Fourier transformation, the solution $x$ is given by
    \begin{align}
        \tilde{x}\left(\omega\right)=A\left(\omega;\omega_{0},Q\right)e^{ia\left(\omega;\omega_{0},Q\right)}\tilde{f}\left(\omega\right)/k,
    \end{align}
    where $\tilde{x}\left(\omega\right)$ and $\tilde{f}\left(\omega\right)$ are the Fourier transformations of $x\left(t\right)$ and $f\left(t\right)$, respectively. Here, $\omega_{0}=\sqrt{k/m}$ is the resonance frequency, $Q=m\omega_{0}/b$ is the quality factor, and the resonance amplitude $A\left(\omega;\omega_{0},Q\right)$ and the phase $a\left(\omega;\omega_{0},Q\right)$ are defined by
    \begin{align}
        A\left(\omega;\omega_{0},Q\right)&\equiv\left[\left\{1-\left(\frac{\omega}{\omega_{0}}\right)^{2}\right\}^{2}+\left(\frac{\omega}{\omega_{0}Q}\right)^{2}\right]^{-1/2},\label{eq:amplification_factor}
        \\a\left(\omega;\omega_{0},Q\right)&\equiv\Arg\left[\omega_{0}^{2}-\omega^{2}-i\frac{\omega_{0}\omega}{Q}\right].\label{eq:amplification_phase}
    \end{align}
    Note that $A\left(\omega;\omega_{0},Q\right)$ is dimensionless, and its maximum value is $Q$ at $\omega=\omega_{0}$. The EM resonance in the cavity will also have the resonance amplitudes identical to $A\left(\omega;\omega_{0},Q\right)$ with their own $\omega_{0}$ and $Q$.

\section{Perturbations}
    
    \subsection{Covariant perturbation theory}
    \label{sec:covariant_perturbation}
    
    Let us delve into covariant perturbation theory to facilitate a clear and concise discussion. Consider a foliation $\mathcal{F}$ by a one-parameter family of perturbed spacetime $\mathcal{M}_{\epsilon}$, where $\epsilon$ is the dimensionless perturbation parameter. We set $\mathcal{M}_{0}$ as the background spacetime. To discuss perturbations, we introduce a one-parameter group of diffeomorphisms $\phi_{\epsilon}:\mathcal{M}_{0}\rightarrow\mathcal{M}_{\epsilon}$. Then, the perturbed quantity for a tensor $X$ is given by the pullback through $\phi_{\epsilon}$ as $\tilde{X}\left(\epsilon\right)\equiv\phi_{-\epsilon}^{*}X$. When $\tilde{X}\left(\epsilon\right)=O\left(\epsilon^{n}\right)$ for a positive integer $n$, it is convenient to introduce a quantity $Y$ such that $\tilde{X}\left(\epsilon\right)=\epsilon^{n}\tilde{Y}\left(\epsilon\right)$, where $\tilde{Y}\left(\epsilon\right)\equiv\phi_{-\epsilon}^{*}Y=O\left(1\right)$. Then, the perturbed quantity is expanded by
    \begin{align}
        \tilde{X}\left(\epsilon\right)=\epsilon^{n}\left\{Y^{\left(0\right)}+\epsilon\left(\delta Y\right)+O\left(\epsilon^{2}\right)\right\},\label{eq:perturbation_expanded}
    \end{align}
    where $Y^{\left(0\right)}=\left.Y\right|_{\epsilon=0}$ is the leading-order value, and $\delta Y$ is the linear perturbation. The linear perturbation is provided by the Lie derivative of the quantity, i.e., $\delta Y=\left[\mathcal{L}_{\upsilon}Y\right]_{\epsilon=0}$, where $\upsilon$ is the five-dimensional vector field on $\mathcal{F}$ generating $\phi_{\epsilon}$. Refer to Fig. 1 in \cite{parkObservationGravitationalWaves2022} for a helpful visualization of the concept.
    
    It is crucial to recognize that the perturbed quantity depends on our choice of $\phi_{\epsilon}$. This introduces mathematical redundancies, or gauges, in the perturbation. The gauge transformation between $\delta Y=\left[\mathcal{L}_{\upsilon}Y\right]_{\epsilon=0}$ and $\delta' Y=\left[\mathcal{L}_{\upsilon'}Y\right]_{\epsilon=0}$, where $\upsilon$ and $\upsilon'$ are generators of $\phi_{\epsilon}$ and $\phi'_{\epsilon}$, respectively, is given by
    \begin{align}
        \delta' Y - \delta Y=\left[\mathcal{L}_{\xi}Y\right]_{\epsilon=0},
    \end{align}
    where $\xi\equiv{}\upsilon'-\upsilon$. Moreover, $\left.\xi\right|_{\epsilon=0}$ is tangent to $\mathcal{M}_{0}$ because $\xi\left(\epsilon\right)=\upsilon'\left(\epsilon\right)-\upsilon\left(\epsilon\right)=1-1=0$, where $\epsilon$ is understood as the scalar field on $\mathcal{F}$. Hence, we can evaluate $\left[\mathcal{L}_{\xi}Y\right]_{\epsilon=0}$ using only quantities in $\mathcal{M}_{0}$, which means that $\left[\mathcal{L}_{\xi}Y\right]_{\epsilon=0}=\mathcal{L}_{\left.\xi\right|_{\epsilon=0}}\left[Y^{\left(0\right)}\right]$. By the formulation of $\left.\xi\right|_{\epsilon=0}$, we can generate all possible linear perturbations using the gauge transformation.
    
    In the process of determining $\delta Y$ from a measurement, the background spacetime $\mathcal{M}_{0}$ and the leading-order value $Y^{\left(0\right)}$ in $\mathcal{M}_{0}$ are typically known, and the experimental measurement $Y$ is performed in the perturbed spacetime $\mathcal{M}_{\epsilon}$. To determine $\delta Y$ from \cref{eq:perturbation_expanded}, ignoring higher-order terms, one needs to choose a gauge to specify the perturbed value $\tilde{Y}\left(\epsilon\right)$. Owing to the absence of a preferred gauge, $\delta Y$ cannot be uniquely determined when it is gauge dependent, making it not measurable. Therefore, gauge invariance for the linear perturbation is essential to enable its measurement. According to the lemma from \cite{stewart1974}, $\delta Y$ is gauge invariant if and only if $Y^{\left(0\right)}$ is zero, a constant scalar, or constructed by the Kronecker delta with constant coefficients.
    
    Let us consider perturbations of spacetime quantities. The perturbed metric expanded as
    \begin{align}
        \tilde{g}_{ab}\left(\epsilon\right)=g^{\left(0\right)}_{ab}+\epsilon h_{ab}+O\left(\epsilon^{2}\right),
    \end{align}
    where $g^{\left(0\right)}$ is the leading-order metric, and $h\equiv\delta g$ is the linear perturbation. The perturbation of the covariant derivative with the Levi-Civita connection $\nabla$ associated with $g$ for a rank (1,1) tensor $X$ is given by
    \begin{multline}
        \delta\left(\nabla_{c}\tensor{X}{^{a}_{b}}\right)=\nabla^{\left(0\right)}_{c}\tensor{\left(\delta X\right)}{^{a}_{b}}+\tensor{(X^{\left(0\right)})}{^{d}_{b}}\tensor{\left(\delta C\right)}{^{a}_{dc}}
        \\-\tensor{(X^{\left(0\right)})}{^{a}_{d}}\tensor{\left(\delta C\right)}{^{d}_{bc}},
    \end{multline}
    where $\nabla^{\left(0\right)}$ is the Levi-Civita connections associated with $g^{\left(0\right)}$ on $\mathcal{M}_{0}$, $X^{\left(0\right)}$ is the leading-order, $\delta X$ is the linear perturbation, and 
    \begin{align}
        \tensor{\left(\delta C\right)}{^{a}_{bc}}&=\frac{1}{2}g^{ad}_{\left(0\right)}\left(\nabla^{\left(0\right)}_{c}h_{bd}+\nabla^{\left(0\right)}_{b}h_{cd}-\nabla^{\left(0\right)}_{d}h_{bc}\right),
    \end{align}
    where $g^{ab}_{\left(0\right)}$ is the inverse of $g^{\left(0\right)}_{ab}$. The Ricci identity for a vector $X$ is given by
    \begin{align}
        \tensor{R}{^{a}_{bcd}}X^{b}=2\nabla_{[c}\nabla_{d]}X^{a},
    \end{align}
    where $\tensor{R}{^{a}_{bcd}}$ is the Riemann tensor. Introducing perturbations to both sides, we obtain the perturbation of the Riemann tensor as
    \begin{align}
        g^{be}\tensor{\left(\delta R\right)}{^{a}_{ecd}}=-2\nabla^{\left(0\right)}_{[c}\nabla_{\left(0\right)}^{[a}\tensor{h}{_{d]}^{b]}}-h^{e[a}\tensor{(R^{\left(0\right)})}{^{b]}_{ecd}},\label{eq:perturb_Riemann_tensor}
    \end{align}
    where $\tensor{(R^{\left(0\right)})}{^{a}_{bcd}}$ is the leading-order Riemann tensor.
    
    So far, we have not introduced any coordinate system in the development of perturbation theory, nor do we need it. However, for readers familiar with perturbations using coordinate systems, we provide the perturbation of a coordinate system and its relation to our approach. Given the coordinate system $\{x^{\alpha}_{\left(0\right)}\}$ on $\mathcal{M}_{0}$, we introduce the adapted coordinate system $\left\{x^{\alpha}\right\}$ on $\mathcal{M}_{\epsilon}$ corresponding to a gauge $\phi_{\epsilon}$ such that $x^{\alpha}=x^{\alpha}_{\left(0\right)}\cdot\phi_{-\epsilon}$. The perturbation of the adapted coordinate system is then given by
    \begin{align}
        \delta x^{\alpha}=\mathcal{L}_{\upsilon}\left(x^{\alpha}\right)=0,
    \end{align}
    where $\upsilon$ is the generating vector field for $\phi_{\epsilon}$. Utilizing the commutativity of the Lie derivative and exterior derivative for the scalar field, we find the perturbation of coordinate dual basis $\left\{dx^{\alpha}\right\}$ as
    \begin{align}
        \delta\left(\left(dx^{\alpha}\right)_{a}\right)=0.
    \end{align}
    Using the fact $\left(\partial\middle/\partial x^{\beta}\right)^{a}\left(dx^{\alpha}\right)_{a}=\tensor{\delta}{^{\alpha}_{\beta}}$, we obtain the perturbation of the coordinate basis $\left\{\partial\middle/\partial x^{\alpha}\right\}$ as
    \begin{align}
        \delta\left(\left(\partial\middle/\partial x^{\alpha}\right)^{a}\right)=0.\label{eq:perturb_of_partial}
    \end{align}
    Finally, we express the perturbation of tensor components as
    \begin{align}
        \delta\left(\tensor{X}{^{\alpha}_{\beta}}\right)=\delta\left(\tensor{X}{^{a}_{b}}\left(dx^{\alpha}\right)_{a}\left(\partial\middle/\partial x^{\beta}\right)_{b}\right)=\tensor{\left(\delta X\right)}{^{\alpha}_{\beta}},\label{eq:perturbation_of_components}
    \end{align}
    where $X$ is a rank (1,1) tensor on $\mathcal{M}_{\epsilon}$, and the components of $\delta X$ are evaluated with the coordinate system $\{x_{\left(0\right)}^{\alpha}\}$ on $\mathcal{M}_{0}$.
    
    To obtain the transformation of adapted coordinate systems, let us consider coordinate systems $\left\{x^{\alpha}\right\}$ and $\left\{x^{\prime\alpha}\right\}$ in $\mathcal{M}_{\epsilon}$ adapted to $\phi_{\epsilon}$ and $\phi'_{\epsilon}$, respectively. The perturbation of their difference in $\phi_{\epsilon}$ becomes
    \begin{align}
        \delta\left(x^{\prime\alpha}-x^{\alpha}\right)=\delta x^{\prime\alpha}=-\left(\delta'-\delta\right)x^{\prime\alpha}=-\mathcal{L}_{\xi}x_{\left(0\right)}^{\alpha}.
    \end{align}
    Introducing pullbacks $\tilde{x}^{\alpha}\left(\epsilon\right)$ and $\tilde{x}^{\prime\alpha}\left(\epsilon\right)$ for $\left\{x^{\alpha}\right\}$ and $\left\{x^{\prime\alpha}\right\}$, respectively, by $\phi_{\epsilon}$, we get the relation given by
    \begin{align}
        \tilde{x}^{\prime\alpha}\left(\epsilon\right)-\tilde{x}^{\alpha}\left(\epsilon\right)=-\epsilon\xi^{\alpha}+O\left(\epsilon^{2}\right).
    \end{align}
    Many approaches to perturbation theory often start from this relation, but in our approach, it is a consequence.
    
    \subsection{Perturbation of Minkowski spacetime}
    \label{sec:perturbation_Minkowski}
    
    Henceforth, we assume that all quantities exist in $\mathcal{M}_{0}$ unless explicitly stated. We also omit the superscript $\left(0\right)$ for brevity when referring to leading-order quantities. Defining $\mathcal{M}_{0}$ as Minkowski spacetime, we have the flat metric $g$, its associated Levi-Civita connection $\nabla$, and the vanishing Riemann tensor $R=0$ at the leading order. From \cref{eq:perturb_Riemann_tensor}, the linear perturbation of the Riemann tensor is then given by
    \begin{align}
        \tensor{\left(\delta R\right)}{^{ab}_{cd}}=-2\nabla^{[a}\nabla_{[c}\tensor{h}{^{b]}_{d]}},\label{eq:Riemann_tensor}
    \end{align}
    which is gauge invariant due to its vanishing leading order. Utilizing the commutativity of perturbation and self-contraction, we derive the perturbation of the Ricci tensor as
    \begin{align}
        \delta\left(\tensor{R}{^{c}_{acb}}\right)=-\frac{1}{2}\Box h_{ab}-\frac{1}{2}\nabla_{b}\nabla_{a}\tensor{h}{^{c}_{c}}+\nabla_{(a}\nabla^{c}h_{b)c}.
    \end{align}
    The perturbation of the Ricci scalar is expressed as
    \begin{align}
        \delta\left(\tensor{R}{^{c}_{acb}}g^{ab}\right)=\nabla_{b}\nabla_{a}h^{ab}-\Box\tensor{h}{^{a}_{a}}.
    \end{align}
    Assuming the perturbed stress--energy tensor as $\tilde{T}\left(\epsilon\right)=O\left(\epsilon^{2}\right)$, the perturbation of Einstein's equations in \cref{eq:Einsteins_equations} is given by
    \begin{multline}
        0=-\frac{1}{2}\Box h_{ab}-\frac{1}{2}\nabla_{b}\nabla_{a}\tensor{h}{^{c}_{c}}+\nabla_{(a}\nabla^{c}h_{b)c}
        \\-\frac{1}{2}g_{ab}\left(\nabla_{d}\nabla_{c}h^{cd}-\Box\tensor{h}{^{c}_{c}}\right).\label{eq:wave_equation_h}
    \end{multline}
    Taking the trace of the above equation reveals the vanishing perturbation of the Ricci scalar, leading to the conclusion that the perturbation of the Ricci tensor also vanishes.
    
    Applying the D'Alembertian to \cref{eq:Riemann_tensor} and utilizing Einstein's equations from \cref{eq:wave_equation_h}, we obtain the wave equation for the perturbation of the Riemann tensor as
    \begin{align}
        \Box\tensor{\left(\delta R\right)}{^{a}_{bcd}}=0.
    \end{align}
    Its wave solution, representing GWs, is given by
    \begin{align}
        \tensor{\left(\delta R\right)}{^{a}_{bcd}}\left(t,\vec{x}\right)=\int d^{2}\kappa\int_{-\infty}^{\infty}\frac{d\omega}{2\pi}\,\tensor{\tilde{R}}{^{a}_{bcd}}\left(\omega,\kappa\right)e^{iP\left(t,\vec{x};\omega,\kappa\right)},
    \end{align}
    where $\left\{t,\vec{x}\right\}$ denotes the globally inertial coordinate system defined in \cref{sec:em_cavity}, $\omega$ is the parameter for angular frequency, $\kappa$ is the unit spatial vector for the propagation direction, $\int d^{2}\kappa$ represents integration over all directions, $\tilde{R}$ is the amplitude, and $P\left(t,\vec{x};\omega,\kappa\right)\equiv\omega\left(-t+\kappa\cdot\vec{x}\right)$ is the phase. Subsequently, \cref{eq:Riemann_tensor} has the general solution, composed of the particular and homogeneous solutions:
    \begin{multline}
        h_{ab}\left(t,\vec{x}\right)=\int d^{2}\kappa\int_{-\infty}^{\infty}\frac{d\omega}{2\pi}\,(\tilde{h}^{\mathrm{p}})_{ab}\left(\omega,\kappa\right)e^{iP\left(t,\vec{x};\omega,\kappa\right)}
        \\+(h^{\mathrm{h}})_{ab}\left(t,\vec{x}\right),
    \end{multline}
    where $h^{\mathrm{h}}\left(t,\vec{x}\right)$ is the homogeneous solution, and $\tilde{h}^{\mathrm{p}}\left(\omega,\kappa\right)$ is the amplitude of the particular solution satisfying
    \begin{align}
        \tensor{\tilde{R}}{^{ab}_{cd}}=2k^{[a}k_{[c}\tensor{(\tilde{h}^{\mathrm{p}})}{^{b]}_{d]}},\label{eq:Riemann_tensor_amplitude}
    \end{align}
    where $k^{a}\equiv\nabla^{a}P$.
    
    For the geodesic congruence, its unit vector field and extrinsic curvature are expanded as follows:
    \begin{align}
        \tilde{u}^{a}\left(\epsilon\right)&=u^{a}+\epsilon\left(\delta u\right)^{a}+O\left(\epsilon^{2}\right),
        \\\tilde{K}_{ab}\left(\epsilon\right)&=K_{ab}+\epsilon\left(\delta K\right)_{ab}+O\left(\epsilon^{2}\right),
    \end{align}
    where the leading order $u$ is the constant four-velocity aligned to the laboratory, $\delta u$ is its linear perturbation, the leading order $K$ vanishes, and $\delta K$ is its linear perturbation. The normalization condition \cref{eq:normalization_u} provides
    \begin{align}
        u\cdot\left(\delta u\right)=-\frac{1}{2}h\left(u,u\right).\label{eq:u_du}
    \end{align}
    Perturbations of \cref{eq:K_definition,eq:K_derivative} give
    \begin{gather}
        \left(\delta K\right)_{ab}=\nabla_{b}\left(\delta u\right)_{a}+u^{c}\tensor{\left(\delta C\right)}{_{acb}},\label{eq:K_perturb}
        \\\nabla_{c}\left(\delta K\right)_{ab}=\nabla_{b}\left(\delta K\right)_{ac}-u_{d}\tensor{\left(\delta R\right)}{^{d}_{acb}}.\label{eq:K_deriv_perturb}
    \end{gather}
    Notice that $\delta K$ is spatial to $u$ and gauge invariant because its leading order vanishes. Contracting indices in \cref{eq:K_deriv_perturb}, we obtain
    \begin{align}
        D^{b}\left(\delta K\right)_{ab}&=0,\label{eq:div_K}
        \\\nabla_{a}\tensor{\left(\delta K\right)}{^{b}_{b}}&=0,\label{eq:deriv_K_trace}
    \end{align}
    where $D$ is the spatial derivative operator defined as $DX=\gamma\left(\nabla X\right)$ for a spatial tensor $X$.

    \subsection{Perturbation of elasticity}
    \label{sec:perturbation_elasticity}

    Perturbed geometrical quantities for elastic material are expressed as follows:
    \begin{align}
        \tilde{v}^{a}\left(\epsilon\right)&=v^{a}+\epsilon\left(\delta v\right)^{a}+O\left(\epsilon^{2}\right),
        \\\tilde{\chi}_{ab}\left(\epsilon\right)&=\chi_{ab}+\epsilon\left(\delta\chi\right)_{ab}+O\left(\epsilon^{2}\right).
    \end{align}
    Considering static material at the leading order, we set $v^{a}=u^{a}$ and $\chi_{ab}=\gamma_{ab}$. Perturbations of the normalization condition \cref{eq:normalization_v}, the Lorentz factor \cref{eq:Lorentz_factor}, and the spatial velocity \cref{eq:spatial_velocity} provide the following:
    \begin{gather}
        u\cdot\left(\delta v\right)=-\frac{1}{2}h\left(u,u\right),
        \\\delta\Gamma=0,
        \\\left(\delta V\right)^{a}=\left(\gamma\delta v\right)^{a}-\left(\gamma\delta u\right)^{a}.\label{eq:delta_V}
    \end{gather}
    \cref{eq:chi_decomposition,eq:chi_symmetry} for the material metric give
    \begin{gather}
        \left(\delta\chi\right)_{ab}=u_{a}\left(\delta V\right)_{b}+\left(\delta V\right)_{a}u_{b}+\left(\delta W\right)_{ab}+\left(\delta\gamma\right)_{ab},
        \\\nabla_{u}\left(\delta W\right)_{ab}=-2D_{(a}\left(\delta V\right)_{b)}-2\left(\delta K\right)_{ab},\label{eq:del_W_evol}
    \end{gather}
    where $\nabla_{u}$ is the covariant derivative along $u$. Perturbations of the spatial metric \cref{eq:gamma_prime} and the strain \cref{eq:strain} yield
    \begin{align}
        \left(\delta e\right)_{ab}&=-\frac{1}{2}\left(\delta W\right)_{ab}.
    \end{align}
    Note that $\delta V$, $\delta W$, and $\delta e$ are spatial and gauge-invariant because their leading-order values vanish.
    
    We introduce the perturbed stress--energy tensor of the material given by
    \begin{align}
        \tilde{T}_{ab}\left(\epsilon\right)&=\epsilon^{2}\left\{T_{ab}+\epsilon\left(\delta T\right)_{ab}+O\left(\epsilon^{2}\right)\right\},
        \\\tilde{\rho}\left(\epsilon\right)&=\epsilon^{2}\left\{\rho+\epsilon\left(\delta\rho\right)+O\left(\epsilon^{2}\right)\right\},
        \\\tilde{\sigma}_{ab}\left(\epsilon\right)&=\epsilon^{2}\left\{\sigma_{ab}+\epsilon\left(\delta\sigma\right)_{ab}+O\left(\epsilon^{2}\right)\right\}.
    \end{align}
    We choose $\tilde{T}\left(\epsilon\right)=O\left(\epsilon^{2}\right)$ to be compatible with the perturbed Einstein equation in \cref{eq:wave_equation_h} and assume that there is no other contribution except the material on $\tilde{T}\left(\epsilon\right)$ up to $\epsilon^{3}$. Considering homogeneous material without stress at the leading order, we set $\nabla_{a}\rho=0$, and $\sigma_{ab}=0$. Then, \cref{eq:material_stress_energy} becomes
    \begin{align}
        T_{ab}&=\rho u_{a}u_{b},
        \\\left(\delta T\right)_{ab}&=\left(\delta \rho\right)u_{a}u_{b}+2\rho\left(u_{(a}\left(\delta v\right)_{b)}+u_{(a}h_{b)c}u^{c}\right)
        \nonumber\\&\qquad\qquad\qquad\qquad\qquad\qquad\qquad+\left(\delta \sigma\right)_{ab}.
    \end{align}
    Note that $\delta\sigma$ is spatial and gauge invariant because its leading order vanishes.

    To derive evolution equations for the material, we perturb the conservation equation \cref{eq:rho_conservation} and the contracted Bianchi identity \cref{eq:contracted_Bianchi_identities} as follows:
    \begin{align}
        \nabla_{u}\left(\delta \rho\right)&=-\rho D\cdot\left(\delta V\right),
        \\\nabla_{u}\left(\delta V\right)^{a}&=-\frac{1}{2}\frac{\lambda}{\rho}D^{a}\tensor{\left(\delta W\right)}{^{b}_{b}}-\frac{\mu}{\rho}D_{b}\left(\delta W\right)^{ab},\label{eq:delta_V_evol}
    \end{align}
    where $D\cdot X$ is the divergence for a spatial vector $X$. Differentiating \cref{eq:delta_V_evol} by $\nabla_{u}$, substituting \cref{eq:del_W_evol}, and utilizing \cref{eq:div_K,eq:deriv_K_trace}, we obtain
    \begin{align}
        0&=-\nabla_{u}^{2}\left(\delta V\right)^{a}+\frac{\lambda+\mu}{\rho}D^{a}\left(D\cdot\delta V\right)+\frac{\mu}{\rho}\Delta\left(\delta V\right)^{a}.\label{eq:eq_for_V}
    \end{align}
    Note that this equation is expressed solely in gauge-invariant quantities, and there is no GW contribution in the above.
    
    The Helmholtz decomposition allows the separation of solenoidal and irrotational modes for $\delta V$. The solenoidal mode (S wave), satisfying $D\cdot\left(\delta V\right)_{\mathrm{S}}=0$, and the irrotational mode (P wave), satisfying $D\times\left(\delta V\right)_{\mathrm{P}}=0$, where $D\times$ is the curl, have wave equations, respectively, as follows:
    \begin{align}
        0&=-\nabla_{u}^{2}\left(\delta V\right)_{\mathrm{S}}^{a}+\frac{\mu}{\rho}\Delta\left(\delta V\right)_{\mathrm{S}}^{a},
        \\0&=-\nabla_{u}^{2}\left(\delta V\right)_{\mathrm{P}}^{a}+\frac{\lambda+2\mu}{\rho}\Delta\left(\delta V\right)_{\mathrm{P}}^{a},
    \end{align}
    where $\nabla_{u}^{2}$ is the second-order time derivative. Similarly, one can derive inhomogeneous wave equations for $\delta W$ that have contributions from GWs.

    \subsection{Perturbation of EM field}
    \label{sec:perturbation_EM}
    
    The perturbed EM field and current are expressed as follows:
    \begin{align}
        \tilde{F}_{ab}\left(\epsilon\right)&=\epsilon^{2}\left\{F_{ab}+\epsilon\left(\delta F\right)_{ab}+O\left(\epsilon^{2}\right)\right\},
        \\\tilde{j}_{a}\left(\epsilon\right)&=\epsilon^{2}\left\{j_{a}+\epsilon\left(\delta j\right)_{a}+O\left(\epsilon^{2}\right)\right\}.
    \end{align}
    Because we imposed $\tilde{F}\left(\epsilon\right)=O\left(\epsilon^{2}\right)$, the EM contribution on $\tilde{T}\left(\epsilon\right)$ is at $O\left(\epsilon^{4}\right)$. It is compatible with the perturbed Einstein's equations \cref{eq:wave_equation_h} and the perturbed contracted Bianchi identities \cref{eq:delta_V_evol}. For the leading orders, we set $j_{a}=0$ and $F_{ab}=\tensor{\varepsilon}{^{c}_{ab}}B_{c}$ where $B$ is the constant magnetic field. Notice that $\delta j$ is gauge invariant because its leading order vanishes. Then, the linear perturbation of \cref{eq:Maxwells_eq_1,eq:Maxwells_eq_2,eq:Maxwells_eq_in_wave_eq_form} becomes:
    \begin{align}
        \nabla^{b}\left(\delta F\right)_{ab}&=4\pi\left(\delta j\right)_{a}-2F_{c[a}\nabla^{b}\tensor{h}{^{c}_{b]}}-\frac{1}{2}F_{ab}\nabla^{b}\tensor{h}{^{c}_{c}},\label{eq:divergence_F}
        \\d\left(\delta F\right)&=0,\label{eq:d_delta_F}
    \end{align}
    \begin{multline}
        \Box\left(\delta F\right)_{ab}+2F_{d[a}\nabla^{c}\tensor{\left(\delta C\right)}{^{d}_{b]c}}
        \\=-4\pi\left(d\left(\delta j\right)\right)_{ab}-F_{cd}\tensor{\left(\delta R\right)}{^{cd}_{ab}}.\label{eq:wave_eq_delta_F}
    \end{multline}
        
    From the definition of the electric field in \cref{eq:E_field}, we obtain its linear perturbation as follows:
    \begin{align}
        \left(\delta E\right)_{a}&=\left(\delta F\right)_{ab}u^{b}+F_{ab}\left(\delta v\right)^{b}.
    \end{align}
    Note that $\delta E$ is spatial and gauge invariant because its leading order vanishes. By contracting $u$ with \cref{eq:divergence_F,eq:wave_eq_delta_F}, we obtain equations for $\delta E$ as
    \begin{align}
        D^{a}\left[\left(\delta E\right)_{a}-F_{ab}\left(\delta V\right)^{b}\right]&=-4\pi\left(u\cdot \left(\delta j\right)\right),\label{eq:div_delta_E}
        \\\Box\left[\left(\delta E\right)_{a}-F_{ab}\left(\delta V\right)^{b}\right]&=D_{a}\left[D^{b}\left(\left(\delta E\right)_{b}-F_{bc}\left(\delta V\right)^{c}\right)\right]
        \nonumber\\&\qquad+4\pi\nabla_{u}\left(\gamma\delta j\right)_{a}
        \nonumber\\&\qquad-F_{cd}\tensor{\left(\delta R\right)}{^{cd}_{ab}}u^{b}.\label{eq:wave_eq_delta_E}
    \end{align}
    Note that these equations are written in only gauge-invariant quantities.

    \subsection{Inside the cavity}
    \label{sec:inside_cavity}

    To derive the perturbations of conditions \cref{eq:vacuum_condition_1,eq:j_at_cavity} within the cavity, let us introduce $\mathcal{W}_{\epsilon}$ as the four-dimensional volume for the cavity motion in $\mathcal{M}_{\epsilon}$. By considering a gauge $\phi_{\epsilon}$ such that $\phi_{\epsilon}\left[\mathcal{W}_{0}\right]=\mathcal{W}_{\epsilon}$, the perturbations of \cref{eq:vacuum_condition_1,eq:j_at_cavity} on $\mathcal{W}_{0}$ can be expressed as follows:
    \begin{align}
        \left.\left(\delta V\right)^{a}\right|_{\mathcal{W}_{0}}&=0,
        \\\left.\left(\delta j\right)_{a}\right|_{\mathcal{W}_{0}}&=0.
    \end{align}
    Because $\delta V$ and $\delta j$ are gauge invariant, these conditions hold in any gauge. Subsequently, within the cavity, \cref{eq:div_delta_E,eq:wave_eq_delta_E} transform into the following equations:
    \begin{align}
        \left.D\cdot\left(\delta E\right)\right|_{\mathcal{W}_{0}}&=0,\label{eq:div_delta_E_cavity}
        \\\left.\Box\left(\delta E\right)_{a}\right|_{\mathcal{W}_{0}}&=\left.-F_{cd}\tensor{\left(\delta R\right)}{^{cd}_{ab}}u^{b}\right|_{\mathcal{W}_{0}}.\label{eq:box_delta_E_cavity}
    \end{align}
    In a similar discussion, one can prove that the following condition, originating from \cref{eq:boundary_condition_E_p}, remains valid in any gauge at the boundary:
    \begin{align}
        \left.\tensor{P}{^{b}_{a}}\left(\delta E\right)_{b}\right|_{\partial\mathcal{W}_{0}}=0.\label{eq:boundary_condition_delta_E}
    \end{align}

    \subsection{Gauges and rigidity}
    \label{sec:gauges}

    \cref{eq:div_delta_E_cavity,eq:box_delta_E_cavity,eq:boundary_condition_delta_E} hold for all gauges. Moreover, each term in these equations is gauge invariant as they involve only gauge-invariant quantities. The contribution of GWs comes from only the right-hand side in \cref{eq:box_delta_E_cavity}. This term vanishes when the direction of GWs aligns with the magnetic field direction due to \cref{eq:Riemann_tensor_amplitude}. This observation contradicts the claims made in \cite{berlinDetectingHighfrequencyGravitational2022,domckeNovelSearchHighFrequency2022}. Let us discuss the reason.

    In \cite{berlinDetectingHighfrequencyGravitational2022}, they impose rigidity of the material by fixing the four-velocity for conductor elements $v$ as $\left(\partial\middle/\partial t\right)$ in the proper detector frame even when GWs pass. Then, $\delta v=0$ due to \cref{eq:perturb_of_partial} in the proper detector gauge, and \cref{eq:divergence_F,eq:wave_eq_delta_F} become
    \begin{align}
        D\cdot\left(\delta E\right)&=-4\pi\left(u\cdot\left(\delta j\right)\right)-F_{ab}u^{c}\tensor{\left(\delta C\right)}{^{a}_{c}^{b}},\label{eq:Berlin_div_E}
        \\\Box\left(\delta E\right)_{a}&=D_{a}\left(D\cdot\left(\delta E\right)\right)+4\pi\nabla_{u}\left(\gamma\delta j\right)_{a}
        \nonumber\\&\qquad-F_{cd}\tensor{\left(\delta R\right)}{^{cd}_{ab}}u^{b}
        \nonumber\\&\qquad+F_{ab}u^{c}\nabla^{d}\tensor{\left(\delta C\right)}{^{b}_{cd}}+F_{bc}u^{d}D_{a}\tensor{\left(\delta C\right)}{^{b}_{d}^{c}}.\label{eq:Berlin_box_E}
    \end{align}
    In these equations, the second term on the right-hand side in \cref{eq:Berlin_div_E} represents the ``effective charge," while the third, fourth, and fifth terms on the right-hand side in \cref{eq:Berlin_box_E} constitute the temporal derivative of spatial ``effective current," as defined in \cite{berlinDetectingHighfrequencyGravitational2022}.

    We would like to address three points. Firstly, $\left(\partial/\partial t\right)$ in the perturbed spacetime $\mathcal{M}_{\epsilon}$ is not normalized. To accurately obtain the electric field, it is essential to use a normalized vector in \cref{eq:E_field}. Secondly, $\delta v$ is a gauge-dependent quantity due to $\mathcal{L}_{\xi}v\neq0$ in $\mathcal{M}_{0}$. Consequently, $\delta v$ is not controllable experimentally. Enforcing $\delta v=0$ requires a condition imposed by \cref{eq:delta_V}, such as $\left(\delta V\right)^{a}=-\left(\gamma\delta u\right)^{a}$, where $\delta u$ is the gauge-dependent perturbation in the proper detector gauge. This cannot be the solution for \cref{eq:eq_for_V}, which is the acoustic wave equation with its propagating velocity smaller than the speed of light, while $\delta u$ has the phase of GWs. Thirdly, a perfect rigid body does not exist in the framework of general relativity because it violates causality, as discussed in \cite{Giulini2010}.

\section{Gravitational wave signal}
\label{sec:GW_signal}

    \subsection{Electric field inside cavity}

    We observe that the boundary condition for $\delta E$ in \cref{eq:boundary_condition_delta_E} shares an identical form with \cref{eq:boundary_condition_E_p} in Minkowski spacetime. Therefore, as in \cref{eq:E_superposition}, the solution $\delta E$ for \cref{eq:box_delta_E_cavity} is superposed by resonant modes as
    \begin{align}
        \left(\delta E\right)_{a}\left(t,\vec{x}\right)&=\sum_{n}\mathcal{E}_{n}\left(t\right)\mathfrak{e}^{n}_{a}\left(\vec{x}\right),
    \end{align}
    where $\mathcal{E}_{n}\left(t\right)$ is the time-dependent amplitude for mode $n$ having the dimension identical to the electric field. This allows us to transform \cref{eq:box_delta_E_cavity} into an equation for $\mathcal{E}_{n}\left(t\right)$ similar to the forced oscillation equation in \cref{eq:force_oscillation}:
    \begin{align}
        \frac{\mathcal{V}}{\omega_{n}^{2}}\left(\ddot{\mathcal{E}}_{n}+\frac{\omega_{n}}{Q_{n}}\dot{\mathcal{E}}_{n}+\omega_{n}^{2}\mathcal{E}_{n}\right)=f_{n}\left(t\right),\label{eq:force_oscillation_E}
    \end{align}
    where we include the dissipation term with the quality factor $Q_{n}$ following \cite{berlinDetectingHighfrequencyGravitational2022} and $f_{n}\left(t\right)$ is the external ``force" defined by
    \begin{align}
        f_{n}\left(t\right)\equiv\frac{1}{\omega_{n}^{2}}\int_{\mathcal{V}}d\mathcal{V}\,F_{cd}\tensor{\left(\delta R\right)}{^{cd}_{ab}}u^{b}\mathfrak{e}_{n}^{a}\left(\vec{x}\right),\label{eq:f_n}
    \end{align}
    where $\mathfrak{e}^{a}_{n}\left(\vec{x}\right)$ is the metric dual to $\mathfrak{e}^{n}_{a}\left(\vec{x}\right)$.
    Notice that the factor $\mathcal{V}/\omega_{n}^{2}$ in \cref{eq:force_oscillation_E} makes the dimension of ``force" identical to the dimension of the energy over the electric field.
    
    The solution $\mathcal{E}_{n}\left(t\right)$ is given by
    \begin{align}
        \tilde{\mathcal{E}}_{n}\left(\omega\right)&=A\left(\omega;\omega_{n},Q_{n}\right)e^{ia\left(\omega;\omega_{n},Q_{n}\right)}\tilde{f}_{n}\left(\omega\right)/\mathcal{V},\label{eq:cal_E_tilde_n}
        \\\tilde{f}_{n}\left(\omega\right)&=\mathcal{V}\left|B\right|\left(\frac{\omega}{\omega_{n}}\right)^{2}
        \nonumber\\&\qquad\int d^{2}\kappa\,\tilde{h}_{ab}\left(\omega,\kappa\right)\left(\hat{B}\times\kappa\right)^{a}\bar{\mathfrak{e}}_{n}^{b}\left(\omega,\kappa\right)e^{i\omega\kappa\cdot\vec{x}_{0}}.\label{eq:force}
    \end{align}
    Here, $\tilde{\mathcal{E}}_{n}\left(\omega\right)$ and $\tilde{f}_{n}\left(\omega\right)$ are the Fourier transformations of $\mathcal{E}_{n}\left(t\right)$ and $f_{n}\left(t\right)$, respectively. In \cref{eq:cal_E_tilde_n}, $A\left(\omega;\omega_{n},Q_{n}\right)$ and $a\left(\omega;\omega_{n},Q_{n}\right)$ are defined in \cref{eq:amplification_factor,eq:amplification_phase}. In \cref{eq:force}, $\tilde{h}_{ab}$ is defined by
    \begin{align}
        \tilde{h}_{ab}\left(\omega,\kappa\right)\equiv2\omega^{-2}\tilde{R}_{acbd}\left(\omega,\kappa\right)u^{c}u^{d},
    \end{align}
    satisfying
    \begin{align}
        \tensor{\tilde{h}}{^{a}_{a}}\left(\omega,\kappa\right)&=0,
        \\\tilde{h}_{ab}\left(\omega,\kappa\right)u^{b}&=0,
        \\\tilde{h}_{ab}\left(\omega,\kappa\right)\kappa^{b}&=0.
    \end{align}
    Note that $\tilde{h}_{ab}$ is identical to the amplitude of metric perturbation in transverse-traceless gauge; however, we consider it as a derived quantity from the gauge-invarant perturbation of the Riemann tensor as in Eq.~(27.24) in \cite{thorne2017}. Also, $\left|B\right|\equiv\sqrt{B\cdot B}$, $\hat{B}\equiv B/\left|B\right|$, and $\times$ is the cross product with $\varepsilon$. Additionally, $\vec{x}_{0}$ is a position inside $\mathcal{V}$, and $\bar{\mathfrak{e}}_{n}\left(\omega,\kappa\right)$ is defined by
    \begin{align}
        \bar{\mathfrak{e}}^{a}_{n}\left(\omega,\kappa\right)\equiv\frac{1}{\mathcal{V}}\int_{\mathcal{V}}d\mathcal{V}\left(\vec{x}'\right)\,\mathfrak{e}_{n}^{a}\left(\vec{x}'\right)e^{i\omega\kappa\cdot\vec{x}'}.\label{eq:e_bar}
    \end{align}
    
    \subsection{Signal and antenna pattern}

    In the context of GW detection using mode $n$, we introduce the GW signal $h\left(t\right)$, the detector tensor $\tilde{D}^{ab}\left(\omega,\kappa\right)$, and the transfer function $\tilde{T}\left(\omega\right)$ as follows:
    \begin{align}
        \tilde{\mathcal{E}}_{n}\left(\omega\right)&=\tilde{T}\left(\omega\right)\tilde{h}\left(\omega\right),
        \\\tilde{T}\left(\omega\right)&\equiv A\left(\omega;\omega_{n},Q_{n}\right)e^{ia\left(\omega;\omega_{n},Q_{n}\right)}\left(\frac{\omega}{\omega_{n}}\right)^{2}\left|B\right|,
        \\\tilde{h}\left(\omega\right)&\equiv\int d^{2}\kappa\,\tilde{D}^{ab}\left(\omega,\kappa\right)\tilde{h}_{ab}\left(\omega,\kappa\right)e^{i\omega\kappa\cdot\vec{x}_{0}},\label{eq:h_tilde}
        \\\tilde{D}^{ab}\left(\omega,\kappa\right)&\equiv \left(\hat{B}\times\kappa\right)^{a}\bar{\mathfrak{e}}_{n}^{b}\left(\omega,\kappa\right).
    \end{align}
    Here, $\tilde{h}\left(\omega\right)$ represents the Fourier transformation of $h\left(t\right)$, with the note that $h\left(t\right)$ is dimensionless.
    
    We present the expression for $\tilde{h}_{ab}$ following the format in Sec. 7 of \cite{parkObservationGravitationalWaves2022}:
    \begin{multline}
        \tilde{h}_{ab}\left(\omega,\kappa;H,\chi,\psi\right)
        \\=H\left(\cos\chi e^{+}_{ab}\left(\kappa;\psi\right)+i\sin\chi e^{\times}_{ab}\left(\kappa;\psi\right)\right)e^{i\xi}.
    \end{multline}
    Here, $H\left(\omega,\kappa\right)>0$ represents the amplitude strength, $\chi\left(\omega,\kappa\right)$ is the ellipticity, $\psi\left(\omega,\kappa\right)$ is the polarization angle, and $\xi\left(\omega,\kappa\right)$ is the phase. The real orthonormal basis $\left\{e^{+},e^{\times}\right\}$ is defined by:
    \begin{align}
        \begin{bmatrix}
            e^{+}\left(\kappa;\psi\right)
            \\e^{\times}\left(\kappa;\psi\right)
        \end{bmatrix}&=
        \begin{bmatrix}
            \cos\left(2\psi\right)&\sin\left(2\psi\right)
            \\-\sin\left(2\psi\right)&\cos\left(2\psi\right)
        \end{bmatrix}
        \begin{bmatrix}
            e^{+}\left(\kappa;0\right)
            \\e^{\times}\left(\kappa;0\right)
        \end{bmatrix},
        \\e^{+}_{ab}\left(\kappa;0\right)&=\frac{1}{\sqrt{2}}\left(x_{a}x_{b}-y_{a}y_{b}\right),
        \\e^{\times}_{ab}\left(\kappa;0\right)&=\frac{1}{\sqrt{2}}\left(x_{a}y_{b}+y_{a}x_{b}\right).
    \end{align}
    Here, $\left\{x,y\right\}$ belongs to the right-handed orthonormal frame $\left\{u,x,y,\kappa\right\}$. Note that the factor $1/\sqrt{2}$ ensures normalization of the basis, satisfying $e_{ab}^{A}e_{cd}^{B}g^{ac}g^{bd}=\delta^{AB}$, where $A,B\in\left\{+,\times\right\}$.
    
    We rewrite \cref{eq:h_tilde} as
    \begin{align}
        \tilde{h}\left(\omega\right)=\int d^{2}\kappa \,H\left(\omega,\kappa\right)\tilde{F}\left(\omega,\kappa;\chi,\psi\right)e^{i\omega\kappa\cdot\vec{x}_{0}}e^{i\xi\left(\omega,\kappa\right)},
    \end{align}
    where $\tilde{F}\equiv\tilde{D}^{ab}\left(\cos\chi e^{+}_{ab}+i\sin\chi e^{\times}_{ab}\right)$ is the pattern function. If the detector tensor has the form of $\tilde{D}^{ab}=D^{ab}e^{i\xi}$, where $D\left(\omega,\kappa\right)$ is a real tensor and $\xi\left(\omega,\kappa\right)$ is a real phase, we obtain the antenna pattern as
    \begin{align}
        \mathfrak{F}\left(\omega,\kappa\right)&\equiv\sqrt{\frac{1}{2\pi}\int_{0}^{2\pi}d\psi\,\tilde{F}\left(\omega,\kappa;\chi,\psi\right)\tilde{F}^{*}\left(\omega,\kappa;\chi,\psi\right)},
        \nonumber\\&=\sqrt{\frac{1}{2}\Lambda_{abcd}\left(\kappa\right)D^{ab}\left(\omega,\kappa\right)D^{cd}\left(\omega,\kappa\right)},\label{eq:antenna_pattern}
    \end{align}
    where $\Lambda_{abcd}\equiv\sum_{A}e^{A}_{ab}e^{A}_{cd}$. Note that this $\mathfrak{F}\left(\omega,\kappa\right)$ does not depend on the ellipticity $\chi$.
    
    \subsection{Cylindrical cavity}

    For a cylindrical cavity, we observe that $\bar{\mathfrak{e}}_{n}^{a}\left(\omega,\kappa\right)$ defined in \cref{eq:e_bar} has the form of a real vector multiplied by a complex scalar because of the parity symmetry, as in \cref{eq:e_symmetry}. Therefore, the antenna pattern given in \cref{eq:antenna_pattern} is applicable to the cylindrical cavity. We introduce a pair of polar and azimuthal angles $\left(\theta,\phi\right)$ for the GW propagation such that $\kappa\left(\theta,\phi\right)=\sin\theta\cos\phi\hat{x}+\sin\theta\sin\phi\hat{y}+\cos\theta\hat{z}$ and parameter $\alpha$ for the magnetic field direction as $\hat{B}\left(\alpha\right)=\cos\alpha\hat{z}+\sin\alpha\hat{x}$. Then, we get the antenna pattern from \cref{eq:antenna_pattern} as
    \begin{align}
        &\mathfrak{F}^{\mathrm{TM}}_{mnps}\left(\omega,\vartheta,\varphi\right)=4\mathfrak{j}_{mn}\mathfrak{A}^{\mathrm{TM}}_{mnp}\left[1+\left(\frac{R}{L}\frac{p\pi}{\mathfrak{j}_{mn}}\right)^{2}\right]
        \nonumber\\&\qquad\times\left|L\omega\frac{\mathfrak{C}_{\left(-1\right)^{p+1}}\left(L\omega\cos\vartheta/2\right)}{\left(L\omega\cos\vartheta\right)^{2}-\left(p\pi\right)^{2}}\right|
        \nonumber\\&\qquad\times\left|\frac{\mathfrak{J}_{m}\left(R\omega\sin\vartheta\right)}{\left(R\omega\sin\vartheta\right)^{2}-\mathfrak{j}_{mn}^{2}}\cos\left(\vartheta\right)\sin\left(\vartheta\right)\mathfrak{C}_{m,-s}\left(\varphi\right)\right|
        \nonumber\\&\qquad\times\sqrt{1-\left(\cos\varphi\sin\vartheta\sin\alpha+\cos\vartheta\cos\alpha\right)^{2}},\label{eq:antenna_pattern_TM}
    \end{align}
    \begin{align}
        &\mathfrak{F}^{\mathrm{TE}}_{mnps}\left(\omega,\vartheta,\varphi\right)=4\mathfrak{j}'_{mn}\mathfrak{A}^{\mathrm{TE}}_{mnp}\left|p\pi\frac{\mathfrak{C}_{\left(-1\right)^{p+1}}\left(L\omega\cos\vartheta/2\right)}{\left(L\omega\cos\vartheta\right)^{2}-\left(p\pi\right)^{2}}\right|
        \nonumber\\&\qquad\times\left[\left(\frac{m}{\mathfrak{j}'^{2}_{mn}}\frac{\mathfrak{J}_{m}\left(R\omega\sin\vartheta\right)}{R\omega\sin\vartheta}\cos\vartheta\mathfrak{C}_{ms}\left(\varphi\right)\right)^{2}\right.
        \nonumber\\&\qquad\qquad\left.+\left(\frac{\mathfrak{J}'_{m}\left(R\omega\sin\vartheta\right)}{\left(R\omega\sin\vartheta\right)^{2}-\mathfrak{j}'^{2}_{mn}}\mathfrak{C}_{m,-s}\left(\varphi\right)\right)^{2}\right]^{\frac{1}{2}}
        \nonumber\\&\qquad\times\sqrt{1-\left(\cos\varphi\sin\vartheta\sin\alpha+\cos\vartheta\cos\alpha\right)^{2}},\label{eq:antenna_pattern_TE}
    \end{align}
    where $\vartheta\equiv\pi-\theta$ and $\varphi\equiv\pi+\phi$.
    
\begin{figure}
        \includegraphics[width=\columnwidth]{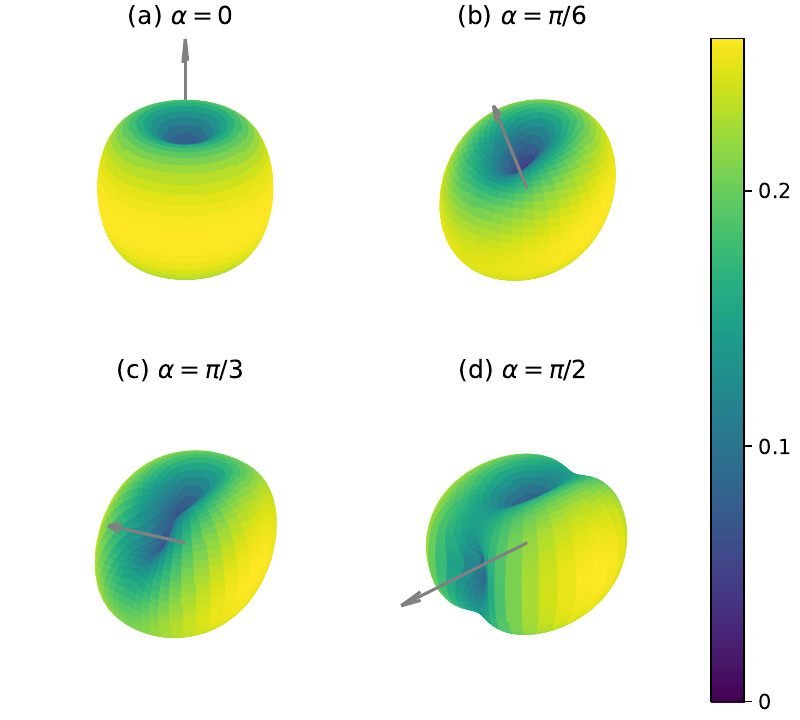}
        \caption{This figure depicts the antenna patterns of cylindrical cavity with $\mathrm{TM}_{010}$ mode, given in \cref{eq:antenna_pattern_TM}, at its resonance frequency. The patterns correspond to magnetic field directions $\alpha\in\left\{0,\pi/6,\pi/3,\pi/2\right\}$ and an aspect ratio $L/R=1$. The gray arrows indicate the magnetic field directions. (a) $\alpha=0$. (b) $\alpha=\pi/6$. (c) $\alpha=\pi/3$. (d) $\alpha=\pi/2$.}
        \label{fig:antenna_TM010}
    \end{figure}

\begin{figure}
        \includegraphics[width=\columnwidth]{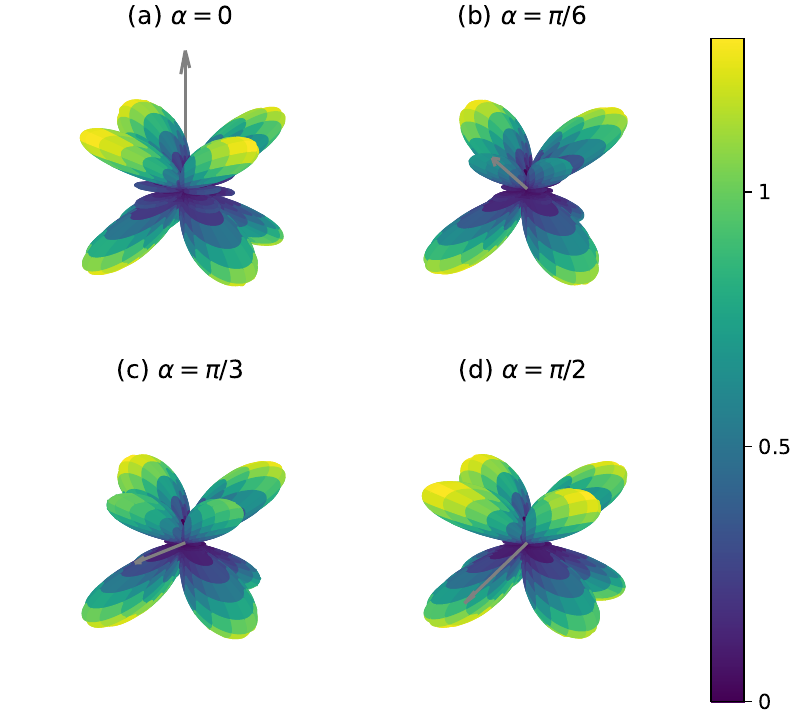}
        \caption{This figure depicts the antenna patterns of cylindrical cavity with $\mathrm{TE}_{212-}$ mode, given in \cref{eq:antenna_pattern_TE}, at its resonance frequency. The patterns correspond to magnetic field directions $\alpha\in\left\{0,\pi/6,\pi/3,\pi/2\right\}$ and an aspect ratio $L/R=1$. The gray arrows indicate the magnetic field directions. (a) $\alpha=0$. (b) $\alpha=\pi/6$. (c) $\alpha=\pi/3$. (d) $\alpha=\pi/2$.}
        \label{fig:antenna_TE212-}
    \end{figure}

\begin{figure}
        \includegraphics[width=\columnwidth]{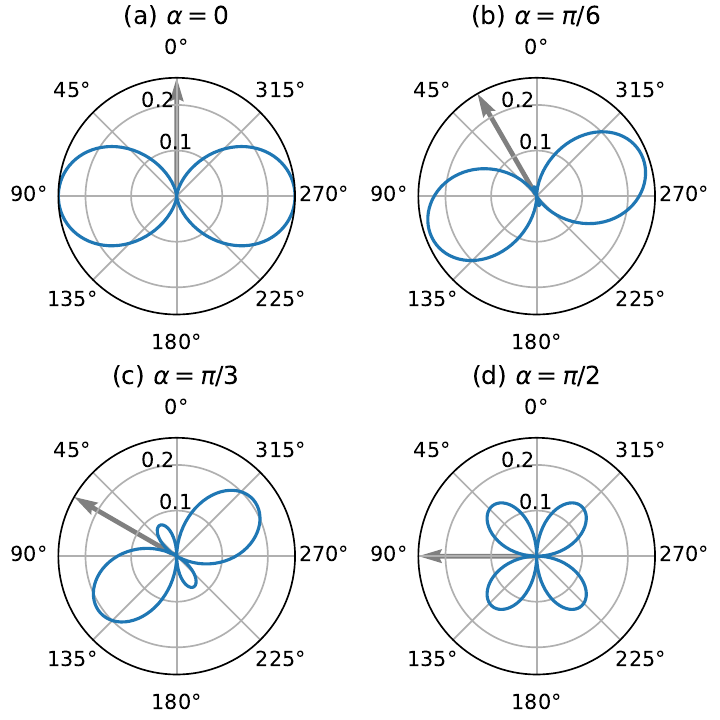}
        \caption{The figure illustrates the antenna patterns of cylindrical cavity with $\mathrm{TM}_{010}$ mode, given in \cref{eq:antenna_pattern_TM}, at its resonance frequency. The patterns are depicted on the plane of $\varphi=0$ and $\varphi=\pi$, corresponding to magnetic field directions $\alpha\in\left\{0,\pi/6,\pi/3,\pi/2\right\}$, with an aspect ratio $L/R=1$. The gray arrows indicate the magnetic field directions. (a) $\alpha=0$. (b) $\alpha=\pi/6$. (c) $\alpha=\pi/3$. (d) $\alpha=\pi/2$.}
        \label{fig:antenna_TM010_2D}
    \end{figure}

\begin{figure}
        \includegraphics[width=\columnwidth]{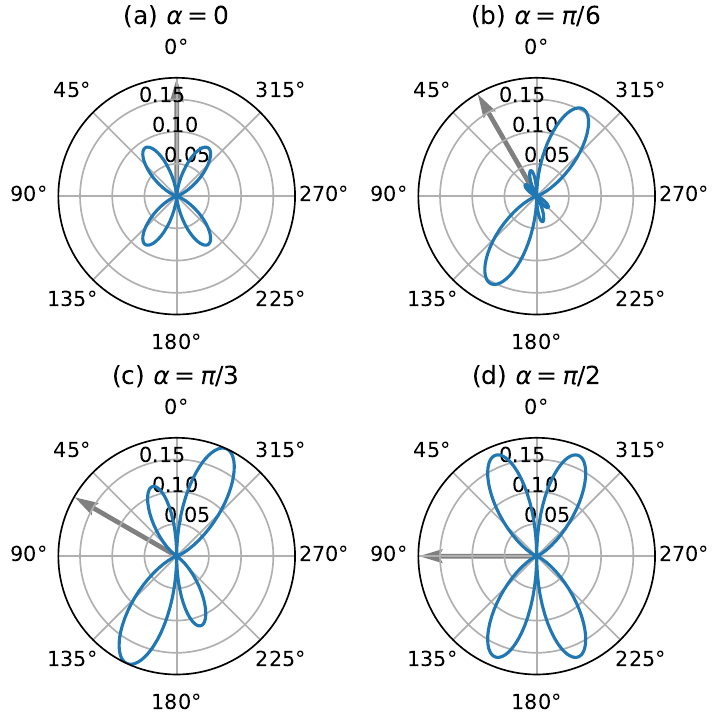}
        \caption{The figure illustrates the antenna patterns of cylindrical cavity with $\mathrm{TE}_{212-}$ mode, given in \cref{eq:antenna_pattern_TE}, at its resonance frequency. The patterns are depicted on the plane of $\varphi=0$ and $\varphi=\pi$, corresponding to magnetic field directions $\alpha\in\left\{0,\pi/6,\pi/3,\pi/2\right\}$, with an aspect ratio $L/R=1$. The gray arrows indicate the magnetic field directions. (a) $\alpha=0$. (b) $\alpha=\pi/6$. (c) $\alpha=\pi/3$. (d) $\alpha=\pi/2$.}
        \label{fig:antenna_TE212-_2D}
    \end{figure}

    \cref{fig:antenna_TM010,fig:antenna_TE212-} show antenna patterns using $\mathrm{TM}_{010}$ and $\mathrm{TE}_{212-}$ modes, respectively, at the resonance frequency for $\alpha\in\left\{0,\pi/6,\pi/3,\pi/2\right\}$ and $L/R=1$. Antenna patterns for $\alpha>\pi/2$ can be obtained by rotating the patterns in the figures by $\pi$ around the axis of cylinder and reversing the direction of the gray arrows. This symmetry arises from the relation $\mathfrak{F}\left(\omega,\vartheta,\varphi;\alpha\right)=\mathfrak{F}\left(\omega,\vartheta,\pi+\varphi;\pi-\alpha\right)$. \cref{fig:antenna_TM010_2D,fig:antenna_TE212-_2D} show cross sections of the antenna patterns on a plane that contains the magnetic field vectors, for $\mathrm{TM}_{010}$ and $\mathrm{TE}_{212-}$ modes, respectively. The panel (a) in \cref{fig:antenna_TE212-_2D} distinctly illustrates a difference between our antenna patterns and those presented in \cite{berlinDetectingHighfrequencyGravitational2022}. Our results reflect the fact that there is no GW signal when the direction of GWs is parallel to the cylinder axis or the magnetic field.
    
\section{Summary and discussion}

We introduce physical laws in a globally hyperbolic spacetime described by Einstein's equations. The contracted Bianchi identities imply the conservation of stress--energy tensor. The congruence of the timelike geodesics without vorticity, have their spatial distribution determined by the extrinsic curvature. However, material elements, despite being influenced by gravity, deviate from geodesics due to elasticity stemming from their deformation. To account for the acoustic oscillation of the material, we introduce the four-velocity and strain, both contributing the stress--energy tensor. The equations of motion for these material elements encompass the temporal symmetry of the material metric, energy conservation, and the conservation of stress--energy tensor. The electromagnetic field is governed by Maxwell's equations. The electric field and magnetic field to the material elements are determined by the four-velocities of the elements. Considering the cavity as a vacuum, we impose the condition of vanishing spatial velocity and EM current. In cases where the material acts as a perfect conductor, the electric field is constrained to be orthogonal to the surface at the boundary.

For a concise discussion, we utilize covariant perturbation theory. Within this framework, we introduce the lemma that a linear perturbation is gauge invariant if and only if the Lie derivative of its leading-order value vanishes along all directions. The perturbation of Einstein's equations transforms into the wave equation for the Riemann tensor, where its wave solutions manifest as GWs. Utilizing these solutions, we present the perturbation of the timelike geodesic congruence. The perturbations of spatial velocity and strain for the material exhibit homogeneous and inhomogeneous wave equations, respectively, as dictated by their equations of motion. Perturbing Maxwell's equations yield the wave equation for the perturbation of the electric field. Within the cavity, this equation indicates that GWs contribute to the perturbation of the electric field solely through the coupling between the EM field and the Riemann tensor. Lastly, we point out that fixing the four-velocity of the material in a coordinate system or considering a rigid body within the context of general relativity is not appropriate.

We solve Maxwell's equations within the cavity. Using the solution for the electric field, we define the transfer function and the dimensionless GW signal. To derive the antenna pattern, we separate the GW signal into its strength and pattern function. Averaging the pattern function over the polarization angle yields the antenna pattern. As illustrative examples, we present the antenna patterns for the EM cavities using the $\mathrm{TM}_{010}$ and $\mathrm{TE}_{212-}$ modes, respectively, at their own resonance frequencies.

Our work enables a profound understanding of the operational principles of GW detectors employing EM cavities by clearly elucidating the interplay among electric fields, acoustic oscillations, and GWs. Owing to the generality of our analysis, it is relevant to various GW detectors employing EM fields. Our paper does not address the measurement principles and methods for the induced EM fields. We perceive this as a challenging problem, and it will be the focus of our future work.

\begin{acknowledgments}
The authors thank Jai-chan Hwang and Sung Mook Lee for their helpful discussion. We appreciate APCTP for its hospitality during the completion of this work. This work was supported by \href{https://ror.org/00hy87220}{IBS} under the Project Codes, No. IBS-R018-D1 and No. IBS-R017-D1-2023-a00. Y.-B.B. was supported by the National Research Foundation of Korea (\href{http://dx.doi.org/10.13039/501100003725}{NRF}) grant funded by the Korean government (\href{http://dx.doi.org/10.13039/501100014188}{MSIT}) (No. NRF-2021R1F1A1051269). C.P. was supported by the \href{http://dx.doi.org/10.13039/501100003725}{NRF} funded by the Korean government (No. NRF-2021M3F7A1082056 and No. NRF-2021R1A2C2012473). All authors contributed equally to this work.
\end{acknowledgments}

\bibliography{ref}

\end{document}